\documentclass[amsmath,twocolumn,superscriptaddress,prx,aps]{revtex4-1} 
\usepackage{amsthm,amsfonts,graphicx,verbatim, color}
\usepackage{bm}
\usepackage[utf8]{inputenc}
\usepackage{ulem}
\let\oldmarginpar\marginpar
\renewcommand\marginpar[1]{\-\oldmarginpar[\raggedleft\footnotesize #1]%
{\raggedright\footnotesize #1}}

\newcommand{\be}{\begin{equation}}
\newcommand{\ee}{\end{equation}}
\newcommand{\bea}{\begin{eqnarray}}
\newcommand{\eea}{\end{eqnarray}}

\renewcommand{\epsilon}{\varepsilon}
\renewcommand{\vec}[1]{{\bf #1}}

\newcommand{\addVR}[1]{{\color{blue}}}

\def\beq{\begin{equation}}
\def\eeq{\end{equation}}
\def\bea{\begin{eqnarray}}
\def\eea{\end{eqnarray}}

\begin{document}

\title{Many body localization with long range interactions}
\author{Rahul M. Nandkishore}
\affiliation{Department of Physics and Center for Theory of Quantum Matter, University of Colorado at Boulder, Boulder CO 80309, USA}
\author{S. L. Sondhi}
\affiliation{Department of Physics, Princeton University, Princeton, NJ  08544, USA}

\begin{abstract}
Many body localization (MBL) has emerged as a powerful paradigm for understanding non-equilibrium quantum dynamics. Folklore based on perturbative arguments holds that MBL only arises in systems with short range interactions. Here we advance non-perturbative arguments indicating that MBL {\it can} arise in systems with long range (Coulomb) interactions. In particular, we show using bosonization that MBL can arise in one dimensional systems with $\sim r$ interactions, a problem that exhibits {\it charge confinement}. We also argue that (through the Anderson-Higgs mechanism) MBL can arise in two dimensional systems with $\log r$ interactions, and speculate that our arguments may even extend to three dimensional systems with $1/r$ interactions. Our arguments are `asymptotic' (i.e. valid up to rare region corrections), yet they open the door to investigation of MBL physics in a wide array of long range interacting systems where such physics was previously believed not to arise.
\end{abstract}
\maketitle
\section{Introduction} 

The phenomenon of many body localization (MBL) has drawn enormous interest from both the theory \cite{Anderson, Gornyi, BAA,Znidaric, OganesyanHuse,Pal,Imbrie, Nandkishore-2015} and experimental \cite{kondov, Schreiber2015,Monroe2016, Choi2016} communities. The intensive investigation of the phenomenon has revealed a cornucopia of exotic physics, including connections to integrability \cite{Bardarson2012, Serbyn, HNO, Scardicchio, lstarbits, GBN, NonFermiGlasses}, unusual response properties \cite{nonlocal, mblconductivity},  a rich pattern of quantum entanglement \cite{Bardarson, Geraedts2016, KhemaniPRX, Chamon, GRN}, and new types of order that cannot arise in equilibrium  \cite{LPQO, Pekkeretal2014, VoskAltman2014}. MBL can also prevent heating in periodically driven `Floquet' systems \cite{DAlessioFloquet, Lazarides, AbaninFloquet, RigolFloquet, KimFloquet} and thus protect new phases of driven quantum matter \cite{KhemaniTC, ElseTC, KeyserTC, LukinTC, MonroeTC,MoessnerFloquet}. MBL has thus emerged as a powerful new paradigm for non-equilibrium quantum dynamics. However, at the same time there has been a proliferation of no go arguments \cite{QHMBL, 2dcontinuum, anycontinuum, proximity, nonabelian, DeRoeck2014, mblmobilityedges, mblbathgeneral, avalanches, SILL} constraining the settings in which MBL can arise. One of the most constraining of these is the restriction \cite{Anderson, LevitovDipoles, Burin, YaoDipoles} to systems with short range interactions. 

The argument against MBL in systems with long range interactions proceeds by examining the convergence of a perturbative `locator expansion' in $d$ dimensional systems when the Hamiltonian contains terms that decay as a power law function of distance, $\sim r^{-\alpha}$. An old argument due to Anderson \cite{Anderson} establishes that hopping terms with $\alpha < d$ lead to breakdown of the locator expansion (see also \cite{LevitovDipoles}). A refinement of this argument \cite{Burin, YaoDipoles} establishes that two body interactions with $\alpha < 2 d$ also `break' the perturbative expansion. Based on these results, a folk theorem has arisen that holds that MBL cannot arise in systems with interactions longer ranged than $1/r^{2d}$. This excludes a great many experimentally relevant systems, including systems of charges (interacting with a Coulomb interaction in any dimension), and systems with dipolar ($1/r^3$) interactions in two and three dimensions. However, this folk theorem rests on shaky foundations, since breakdown of perturbation theory does {\it not} establish breakdown of localization. Intriguingly, recent experiments \cite{LukinTC} with dipoles seem to indicate MBL-type physics in a setting where this folk theorem would suggest such physics cannot arise. Could MBL survive after all in systems with long range interactions? 

In this work we present non-perturbative arguments indicating that MBL {\it can} arise with long range interactions. Our conclusions apply even to interactions longer ranged than $1/r^d$. The key idea is that the long range interactions can drive the system into a non-trivial correlated phase, naturally described in terms of emergent degrees of freedom with only short range interactions. The problem can then be mapped onto the classic analysis of \cite{BAA} to establish many body localization. We demonstrate the viability of this idea, using non-perturbative techniques to treat the interaction, for Coulomb interacting systems in any dimension (i.e. one dimensional systems with $\sim r$ interactions, two dimensional systems with $\log r$ interactions, and three dimensional systems with $1/r$ interactions). The arguments are presented in `decreasing order of rigor,' with the one dimensional analysis being on the firmest footing, and the three dimensional analysis the most speculative. In one dimension our arguments make use of a phase that exhibits charge confinement. In higher dimensions, it makes use of superconductivity and the Anderson-Higgs mechanism. Insofar as our analysis relies on a mapping to \cite{BAA}, it shares similar limitations viz. we can only establish localization at low (but non-zero) temperatures \footnote{Such low temperature MBL is sometimes referred to as `asymptotic' localization \cite{DeRoeck2014}. Asymptotic localization is weaker than `infinite temperature MBL' because it remains possible that rare `hot bubbles' \cite{mblmobilityedges, avalanches} may cause delocalization on exponentially long timescales.  However, even `asymptotic' MBL is more than folk wisdom permits, and indeed is likely to be indistinguishable from `true MBL' \cite{Imbrie} in laboratory experiments.}. Whether infinite temperature localization can arise with long range interactions remains an open problem. However, our work opens the door to the study of MBL physics in a host of experimentally relevant low temperature systems with long range interactions. We emphasize that our work differs from the classic analysis of Ref.\cite{EfrosShklovskii} in that it predicts a strictly zero conductivity at non-zero energy density, up to rare region effects. It differs also from \cite{Kaiser, Santos2} in that we do not restrict ourselves to Anderson localization of single spin flips, and consider instead {\it many body } localization. 

\section{Localization with confining interactions in one dimension}
We begin with a discussion of one dimensional systems, where the long range interaction may be treated exactly using the method of bosonization. 
We start with a continuum model that is inspired by the `Schwinger model' \citep{Schwinger, Coleman, zittarz, Fishler} from high energy physics. The Schwinger model is a model of quantum electrodynamics in one dimension, which exhibits charge confinement. It is formulated as a one dimensional Dirac fermion coupled to a gauge field. For our purposes, however, it is more convenient to adopt a description in which the gauge field has been integrated out. The Hamiltonian that we wish to study thus involves Dirac fermions moving in one dimension with a long range `constant force' interaction (Coulomb interaction in one spatial dimension). 
The Hamiltonian is $H_0 +H_{int}$ where 
\begin{eqnarray}
H_0 &=& \int \frac{dk}{2\pi} \sum_{r=\pm1} v (r k - k_F) c^{\dag}_{r,k} c_{r,k} \label{H_0}; \\
H_{int} &=&  - e^2 \int dx dy \rho(x) |x-y| \rho(y) ; \quad  \label{Hint} 
\end{eqnarray}
and $\rho(x) = \sum_{r=\pm1 } c^{\dag}_{r}(x) c_{r}(x)$. The argument is cleanest when the theory is formulated in the continuum, in which case an asymptotically large UV cutoff $\Lambda$ must be placed on the dispersion. Lattice formulations of the argument will be discussed after we have introduced the main argument. The Schwinger model itself has an additional parameter, which is a uniform background electric field (which can be set to have any value). We choose to set this uniform background electric field to zero. Consideration of potential phase transitions driven by uniform background field is deferred to future work. There is a uniform positively charged `jellium' background (introduced so that the energy density remains finite in the thermodynamic limit). This background is taken to be rigid i.e. we neglect any coupling to `phonons' of the jellium. 

We now make use of standard bosonization formulae from \cite{Giamarchi} to obtain \footnote{To obtain the expression below for $H_{int}$ one can either Fourier transform, bosonize, then inverse Fourier transform, or bosonize and then integrate twice by parts.} 
\begin{eqnarray}
H_0 &=& \int \frac{d x}{(2\pi)} v_F \left[\pi^2 \Pi^2(x) +  (\nabla \phi(x))^2 \right]\\
H_{int} &=& \frac{4 e^2}{\pi} \int \frac{dx}{2\pi} \phi^2(x)
\end{eqnarray}
where $\phi$ describes fluctuations of charge density wave order, and we have introduced $\Pi$ as the conjugate momentum to the $\phi$ field. 
This corresponds to a `Klein Gordon' theory of free massive scalar bosons. The density pattern in the ground state (at a non-zero fermion density) shows crystalline long range order \cite{Fishler}, and the phase may be thus thought of as a kind of `Wigner crystal' where the the usual constraints on long range order in one dimension have been evaded because of the long range interaction. We can make this expression look more familiar if we cast it into the form
\begin{equation}
H = \int \frac{d q}{(2\pi)} u_q  \left[ K_q \pi^2 \Pi(q) \Pi(-q) + \frac{1}{K_q} q^2 \phi(q) \phi(-q)\right] \label{LuttingerLiquid}
\end{equation}
where 
\begin{equation}
u_q = v_F \sqrt{1 + \frac{2 V_q}{\pi v_F}}; \qquad K_q = \frac{1}{\sqrt{1+\frac{2V_q}{\pi v_F}}}; \qquad V_q = \frac{2e^2}{q^2} \label{parameters}
\end{equation}
Adding short range interactions will shift $K(q\rightarrow \infty)$ but will not affect the physics of interest to us here. Note that the bosonized description of the Schwinger model involves a {\it non-compact} boson - this is a consequence of integrating out the gauge field \footnote{We thank Ed Witten for clarifying the origin of the non-compactness}, and will be important to our argument. Note also that so far the transformations performed are {\it exact} at the operator level. 

We now introduce disorder. We emphasize that the `standard prescription' in the localization literature of  perturbing about the infinite disorder state is unsuitable here because of the long range interaction. Not only will the perturbation theory not converge \cite{Burin, YaoDipoles}, but the `infinite disorder' state itself is an illegal starting point - 
a region of size $L$ will have charge $\sim \sqrt{L}$ from central limit theorem, and will thus have an `electrostatic' energy $\sim L^{3/2}$, which will diverge in a super-extensive fashion in the limit of large system size. Instead, we {\it first} treat the interaction exactly by the method of bosonization, and then introduce disorder. Our analysis will be controlled in the regime when disorder is {\it weak} compared to the interaction. We emphasize also that the disorder is allowed to backscatter electrons (turn right movers into left movers and vice versa), which is physics that does not typically enter the high energy literature. The disorder adds to the Hamiltonian a piece $H_{dis}$ where, in accordance with standard formulae \cite{Giamarchi} 
\begin{equation}
H_{dis} = - \int dx\left[ \frac{1}{\pi} \eta(x) \nabla \phi + (\frac{\xi^*(x)}{2\pi \alpha} e^{i 2 \phi(x)} + h.c.) \right]
\end{equation}
here $\eta$ represents forward scattering, and $\xi$ represents backscattering. We make the standard assumption that $\eta$ and $\xi$ can be modeled as independent short range correlated Gaussian random variables. 
Here $\alpha \sim 1/\Lambda$ is a UV cutoff. 

At this stage, it is tempting to apply the classic Giamarchi-Schulz renormalization group analysis \cite{GiamarchiSchulz}, which obtains the $\beta$ functions for the disorder strength $\mathcal{D}$ (defined by $ \langle \xi^*(x) \xi(y) \rangle =\mathcal{ D} \delta(x-y)$), perturbatively in weak disorder. Generalized to the present problem, an analysis of this form gives $\frac{d \mathcal{D}}{dl} = 3 \mathcal{D}$
i.e. disorder is always a relevant perturbation. However the Giamarchi-Schulz calculation is a zero temperature calculation, whereas we are interested in the behavior at non-zero temperatures. Additionally, we wish to consider an {\it isolated} quantum system, disconnected from any external heat bath, such that one cannot use the Matsubara formalism, nor does it make sense to talk about free energy minimization. Finally, we are interested not in the `disorder averaged' properties, but rather in the behavior of a {\it single} sample in a typical disorder realization. 

The analysis of a disordered, interacting Luttinger liquid away from its ground state, without the crutches of disorder averaged field theory or the Matsubara formalism, may appear to be a formidable task  \cite{Mirlinrecent, Mullerrecent, SILL}. It is a problem, however, that is amenable to analytical treatment.
We start by introducing the notation $\xi(x) = D(x) \exp(i \zeta(x))$, and hence rewrite the Hamiltonian (for a particular disorder realization, after an integration by parts) as 
\begin{eqnarray}
H &=& \int \frac{dx}{2\pi} v_F \pi^2 \Pi^2(x) + v_F (\nabla \phi)^2 + V(\phi); \label{Htarget} \\ 
V(\phi) &=&  \frac{4e^2 }{\pi } \phi(x)^2+  \frac{2 D(x)}{ \alpha} \cos(2 \phi - \zeta(x)) - 2 \tilde \eta(x) \phi
\end{eqnarray}
where $\tilde \eta(x) = \int^x dy \eta(y)$, and $\tilde \eta$, $D(x)$ and $\zeta(x)$ are taken to be independent zero-mean random variables with short range correlations. We now introduce $\phi_0(x)$ to be the {\it static} background field configuration that minimizes the Hamiltonian
\begin{equation}
v_F \partial_x^2 \phi_0 =  \frac{4e^2}{\pi}  \phi_0(x) - \frac{2D(x)}{\alpha} \sin (2\phi_0-\zeta) - \tilde \eta
\end{equation}
Note that $\phi_0$ simply represents the adjustment of the Wigner crystal to the disorder that we have introduced i.e. it represents the `classical ground state,' where all quantum fluctuations have been ignored. We assume that $D(x)$ is a {\it bounded} random variable $0 < D(x)< D_0$ with $D_0/\alpha < e^2/\pi$ so that $V(\phi)$ has a {\it unique} minimum. We re-introduce quantum fluctuations by writing $\phi(x) = \phi_0 + \delta \phi$, and obtain the effective Hamiltonian as a power series in small $\delta \phi$. This expansion is well behaved since $V(\phi)$ has a unique minimum, in contrast to the situation that obtains for compact potentials \cite{Giamarchi}, where instantons connecting distinct minima must be taken into account. Relabelling $\delta \phi$ simply as $\phi$, we obtain the Hamiltonian (up to an unimportant additive constant) as 
\begin{eqnarray}
H &=& \int \frac{dx}{2 \pi} v_F \pi^2 \Pi^2(x) + v_F (\nabla \phi)^2 \nonumber\\ &+& \left[ \frac{4 e^2}{\pi} - \frac{4D(x)}{\alpha} \cos(2\phi_0 - \zeta) \right] \phi^2 + O(\phi^3) 
\end{eqnarray}
At leading order this is simply a theory of non-interacting gapped bosons in a random potential, which is well known to have all its (single-particle) eigenstates localized (see e.g. \cite{GurarieChalker2} for an explicit discussion). The higher order terms come from the expansion of the cosine, are short ranged in real space, and may be treated within a perturbative locator expansion. Perturbative locator expansions of this form were shown to converge at sufficiently low (but non-zero) energy densities \cite{BAA} (up to possible rare region corrections \cite{mblmobilityedges}), and thus localization should persist even when non-linearities from higher order terms in the expansion are taken into account i.e. the Hamiltonian (\ref{Htarget}) should exhibit {\it many body localization} at sufficiently low (but non-zero) energy densities. 

There is a subtlety to be noted here. Given that we are working in the continuum, the single particle localization length is unbounded above, whereas \cite{BAA} assumes a bounded localization length. Problems with unbounded single particle localization length and short range interactions have been studied in the MBL literature \cite{aleiner2010finite, 2dcontinuum, mullergornyi}. Ref. \cite{2dcontinuum} showed that as long as $ \Upsilon = g \xi^{3d}(E) P(E)^2 (1-P(E))^2$ is small everywhere in the spectrum (where $g = D/\alpha$ is the non-linearity strength, $\xi(E)$ is the localization length and $P(E)$ is the occupation number for system prepared in a Gibbs state parametrized by a `temperature' $T$), then the locator expansion converges at a typical point in space. Now, at small energy densities (i.e. low `temperatures'), and at small $E$, we surely have $\Upsilon \ll 1$. Meanwhile, since $\xi(E)$ and $g(E)$ both grow as power law functions of energy, but the $P(E)$ function decays as an exponential function of energy, we continue to have $\Upsilon \ll 1$ at high energies, and a mapping onto \cite{BAA} to establish perturbative stability of MBL at a typical point in space is possible \cite{aleiner2010finite}. 

Ref.\cite{2dcontinuum} also raised the possibility that rare `hot' regions may break the locator expansion in the continuum. This scenario was recently studied in detail by \cite{mullergornyi}, who concluded that rare regions would always lead to delocalization, with a relaxation timescale that diverged {\it faster} than $\exp(1/T)$ at low temperatures. This `rare region' problem is endemic to models with many body mobility edges \cite{mblmobilityedges}, and the present model is no exception. Whether the rare region problem can be circumvented remains an open problem, which however has nothing to do with the long range interacting nature of the problem - the Hamiltonian (\ref{Htarget}) is just as localized as would be a short range interacting problem in the continuum. Indeed, the charge confinement in the model makes the interactions between the available degrees of freedom effectively short range (recall that interactions between dipoles in one dimension are not long range), and thus many body localizable in the usual manner, even though the underlying Hamiltonian had long range interactions. 


We now estimate the localization length for the low lying excited states. This is approximately the single particle localization length (since perturbation theory in the interaction converges at typical points in space). In one dimension this is proportional to the scattering length $l$, which can be calculated in self consistent Born approximation (SCBA) with the Green function $G = \frac{1}{E - 4 e^2/\pi -  \hbar v_F k^2 \alpha}$, cutoff on the lengthscale $l$. For states just above the gap this gives $l = (\hbar v_F / D_0 \Xi)^{2/3} \alpha$, where $D_0 \Xi$ is the Fourier transform of the disorder potential (i.e. $\Xi$ is a disorder correlation length), and $\alpha$ is the UV cutoff lengthscale. The analysis is only controlled when $D_0 \Xi \ll \hbar v_F$ i.e. disorder is weak compared to the kinetic energy scale. 

This localization lengthscale should be compared to the `de-Broglie length' $\lambda$, set by the inhomogeneity of the potential. If the disorder is weak (as we are assuming) then the de-Broglie wavelength will be long, and so we should account for central limit averaging of the disorder over one de-Broglie wavelength. We will then have to solve
\begin{equation}
\frac{\hbar v_F }{ \lambda^2 } = \frac{D_0}{\alpha} \sqrt{\Xi/\lambda } \Rightarrow \lambda = \left(\hbar v_F/D_0 \Xi \right)^{2/3} (\alpha^2\Xi)^{1/3} \label{debroglie}
\end{equation}
Note that self consistency requires $\lambda \gg \Xi$, which is automatically ensured at weak disorder. We can now observe that $\frac{l}{\lambda} = \left( \alpha/\Xi \right)^{1/3}$ is the standard control parameter for weak localization theory \cite{AALR}. 
%
When $\Xi > \alpha$ (such that the UV cutoff is the smallest lengthscale in the problem), then $l/\lambda \ll 1$ i.e. the lowest lying excited states are deep in the locator limit. 

\subsection{Lattice regularizations}
Thus far we worked with a model in the continuum. We now discuss lattice regularizations. The natural tight binding lattice generalization of the continuum problem discussed above is 
\begin{equation}
H = \sum_k (E(k) - \mu) c^{\dag}_k c_k - e^2 \sum_{xy} \rho_x \rho_y |x-y| + \sum_x \mu_x \rho_x \label{Hlattice}
\end{equation}
where $\mu_x$ is a random potential, and where $E(k)$ is the bandstructure of the lattice Hamiltonian. We specialize to fermions at {\it incommensurate} filling, leaving the problem of commensurate fillings to future work. Standard phenomenological bosonization \cite{Haldane, Giamarchi, Imambekov} then predicts that the bosonized Hamiltonian will take the form $H_{l} + H_{nl}$, where $H_l$ is (\ref{Htarget}) with integrals replaced by sums and continuum derivatives replaced by lattice derivatives, and $H_{nl}$ contains non-linear corrections (terms of the form $(\nabla \phi)^3$, $(\nabla \phi)^4$ etc) coming from band curvature. At incommensurate filling, when the density can be replaced by the smeared density, the interaction bosonizes to a sum of local terms, and while bosonization does produce non-linear terms, these are strictly short range \cite{Haldane, Imambekov}. Thus the problem still maps (after manipulations analogous to those discussed above in the continuum), to a problem of massive bosons in a random potential with {\it short range} interactions. One can again appeal to \cite{BAA} to argue that this problem should be many body localized. 

We note that by going to a lattice we have eliminated the problem of an `unbounded above' single particle localization length that complicated the analysis in the continuum. However, since the bosonization formulae are only applicable for `almost linear' dispersions, our analysis is still restricted to low (but non-zero) energy densities, when linearization about a Fermi surface is a sensible starting point \footnote{In a sense, we have traded a problem of rare regions containing very high energy excitations, where perturbation theory is inapplicable, for a problem of rare regions containing far from ground state fluctuations, where bosonization is inapplicable. }. Whether {\it infinite temperature} MBL can arise here is an open problem that we leave to future work.

  \section{Localization with Coulomb interactions in higher dimensions}
  
  We now discuss how low temperature MBL may arise in higher dimensional systems with long range interactions. The most natural generalization of our one dimensional example would involve considering Wigner crystals in higher dimensions. However, the distortions of the Wigner crystal interact via dipolar interactions \cite{Fertig}, which in dimensions higher than one are not purely short range. It turns out that if the interaction is sufficiently long range to prohibit dissociation of dipoles, then the interaction between dipoles is itself sufficiently long range to obstruct a locator expansion \cite{Burin}. Conversely, if the interaction between dipoles is sufficiently short range to allow for a locator expansion, then the energy cost of dissociating a dipole is finite, such that at non-zero energy density there exists a non-zero density of free charges, which interact via the `bare' long range interaction. Thus, the obvious generalization of our discussion to higher dimensions is problematic.  
 
 A more fruitful line of attack is opened up by viewing our one dimensional problem as an example of a {\it confining} phase \cite{Schwinger, Coleman}. Given the intimate connections between confinement and the Anderson-Higgs mechanism \cite{FradkinShenker}, we are therefore prompted to consider {\it Higgsed} phases (e.g. superconductors) as a possible platform for higher dimensional MBL with long range interactions. We therefore focus in this section on using {\it superconductivity} to eliminate the long range charge charge interaction, and to obtain a description of a correlated phase in terms of emergent excitations with purely short range interactions, which may then be many body localized. We begin with a discussion in two dimensions, before generalizing to three dimensional systems. The argument works equally well in the continuum or on the lattice, modulo the usual subtleties with localization in the continuum \cite{aleiner2010finite, 2dcontinuum, mullergornyi}. A jellium background is again assumed, so that the uniform state has finite electrostatic energy in the thermodynamic limit. 
  
    It is imperative that we do not have phonons in the problem, since phonons (and Goldstone modes in general) have a diverging single particle localization length at low energies \cite{GurarieChalker} which is believed to pose an obstruction to MBL \cite{QHMBL, BanerjeeAltman}. We thus need a purely electronic mechanism for superconductivity. We use the Kohn Luttinger theorem to this end \cite{KL, KL2}, as a key building block for our analysis. The Kohn Luttinger argument in the continuum \cite{KL, KL2} shows that a long range isotropic repulsion generates through perturbation theory a short range attraction in a sufficiently high angular momentum channel, which can induce superconductivity. Lattice versions of the argument are also known (see e.g. \cite{Raghu, NTC} for recent discussions). That superconductivity arises in a high angular momentum channel is a feature, since these superconductors lack the protection against disorder that s-wave superconductors inherit from the Anderson theorem \cite{AndersonThm}, and are thus easier to localize. However, it is important that the superconductivity should be {\it non-chiral}, since chiral states possess their own obstructions to localization \cite{QHMBL}. This may be accomplished either by working on a lattice where the Kohn Luttinger attraction arises in a one dimensional irreducible representation of the lattice symmetry group (see e.g. \cite{NTC} for a specific example), or in the continuum, if the energetics favor a nodal rather than a chiral state. It is also imperative that there should not be a spin $SU(2)$ symmetry in the problem, since $SU(2)$ symmetry poses its own obstruction to MBL \cite{nonabelian, AbaninSU2}. This may be evaded either by working with spinless fermions, or by applying a Zeeman field to break the spin symmetry down to $U(1)$. 

We now discuss how superconductivity enables MBL in a long range interacting system. The argument is independent of the precise mechanism of superconductivity (as long as it is not mediated by the Goldstone bosons of some continuous symmetry e.g. acoustic phonons), and also of the particular structure of the superconducting ground state - as long as it is non-chiral, not protected by the Anderson theorem, and has low enough symmetry that there are no higher dimensional irreducible representations \cite{nonabelian}. We emphasize that we are discussing here not superfluidity, but rather true superconductivity i.e. the charges are coupled to a dynamical gauge field, and the Goldstone mode is gapped out by the Anderson-Higgs mechanism. We emphasize also that we are discussing a superconductor treated as a {\it closed} quantum system, which is {\it not} in thermodynamic equilibrium. Additionally, the superconductor is {\it disordered}, but the disorder is not so strong as to destroy superconductivity. 

Once the system becomes superconducting, the long range interaction is screened out. The effective degrees of freedom in a superconductor are the Bogolioubov de-Gennes quasiparticles, the vortices, and bound states of the two \cite{HanssonOganesyanSondhi, Moroz}, as well as the photons which mediate the electromagnetic interaction. We emphasize that the correctly formulated excitations carry neither charge nor dipole moment on long lengthscales \cite{Bogolons}. This must be the case, since otherwise there would be electromagnetic fields at long lengthscales, which is inconsistent with Meissner physics. For an s-wave superconductor in two dimensions, the effective theory for quasiparticles and vortices is simply the toric code \cite{HanssonOganesyanSondhi, Moroz}, the topologically ordered phase of which is the superconductor. The disordered toric code has been shown \cite{LPQO} to support topological order even in its {\it excited} states, from which it follows that an isolated two dimensional s-wave superconductor can exhibit superconductivity even away from its ground state, with vortices and quasiparticles localized on disorder. The present problem differs somewhat in that the quasiparticles are nodal rather than gapped. However, at the level of the non-interacting theory, it is known that a two dimensional disordered nodal superconductor supports an Anderson localized phase for the quasiparticles \cite{SF, SFBN, VSF}. Meanwhile, the interactions (between vortices, between quasiparticles, and between quasiparticles and vortices), have been derived in e.g. \cite{VSF}, and are purely short ranged. If we can also demonstrate localization of the photon mode, it will then follow from \cite{BAA} that a system of localized quasiparticles and vortices with weak short range interactions will be in a many body localized phase, notwithstanding that the `bare' electronic Hamiltonian contained a long range interaction. 

We now discuss the localization of the photon mode. 
 In the superconductor the photon mode is gapped out by the Higgs mechanism, and can thus be ignored when ground state physics is the main concern, as in \cite{SF, SFBN, VSF}. However, since we aim to establish MBL at low but non-zero temperatures, the photon mode must be taken into account. We now offer two arguments that the photon mode is also localized, and thus does not materially alter the conclusions reached above. Both arguments are adapted from the equivalent arguments for Goldstone modes in \cite{GurarieChalker}. Note that in the case of the superconductor, the Goldstone mode does not exist as a separate excitation, but instead is `absorbed' into the photon mode via the Anderson-Higgs mechanism. 

We wish to describe a superconductor with order parameter $\Delta(r) \exp(i \theta(r))$ minimally coupled to a gauge field $(A_0, \vec{A})$ which lives in two dimensions. The effective theory for this is the Abelian Higgs model, \cite{HanssonOganesyanSondhi, Moroz}, the equation of motion for which takes the form \cite{wenbook} 
  \begin{equation}
[ \partial_t^2 - c_{L, T} \nabla^2 + \Delta(r)^2 ] \vec{A}_{L,T} = 0
 \label{AbelianHiggs}
  \end{equation}
where $\Delta(r)$ is the (spatially inhomogenous) gap function, $A_{L}$ ($A_T$) is the longitudinal (transverse) photon mode (the longitudinal photon mode being the remnant of the plasma oscillation mode in the normal metal), and $c_{L}$ ($c_T$) is the longitudinal (transverse) photon velocity. In a physical superconductor, $c_T$ is the speed of light while $c_L$ is of order Fermi velocity. If the speed of light is taken to infinity (so that the interaction is instantaneous) then the transverse mode can be neglected as `infinitely energetic,' but the longitudinal polarization must still be taken into account. 
  Note that disorder enters through a mass term, i.e. the disorder vertex does not vanish at low frequency, and the dispersion relation at low frequency takes the form $\omega^2 \approx \Delta(r)^2 + q^2 \Rightarrow \omega = \Delta + q^2/2 \Delta$. We now follow \cite{GurarieChalker} and first estimate a mean free path $l$ from SCBA, and then substitute $k l$ into weak localization theory, where $k$ is the clean system wavevector corresponding to a frequency $\omega$. This analysis reveals that in spatial dimensions $d=1,2, 3$ , $kl$ is free of divergences at low frequency, such that {\it all} low energy photon modes can be localized with bounded localization length, in sharp contrast to (non-Higgsed) Goldstone modes \cite{GurarieChalker} (see Appendix for explicit calculation). Moreover, the interactions between photon modes and order parameter fluctuations are strictly short range, so the photon sector does not present any obstruction to localization.

An alternative argument, also adapted from \cite{GurarieChalker}, proceeds as follows. The dispersion relation for the plasmon mode takes the form 
  \begin{equation}
  \omega^2 = \Delta^2 + q^2 + 2 \Delta m(r) \Rightarrow 2 \Delta  \tilde \omega \sim q^2 + \Delta m(r) \label{plasmondispersion}
  \end{equation}
  where $m$ is the (zero mean) fluctuation in the gap function and $\Delta$ is the mean gap function, and to obtain the second expression we have taken the scaling limit $\omega \rightarrow \Delta$ and have defined $\tilde \omega = \omega - \Delta$. 
 Now performing central limit averaging on the disorder over one wavelength of the clean system we obtain
    \begin{equation}
2 \Delta  \tilde \omega \sim q^2 + \Delta m_0 q^{d/2}
  \end{equation}
  where $m_0$ is the typical fluctuation in the gap function. For $d < 4$ the disorder term dominates the low energy dispersion relation (i.e. disorder is relevant). 
  One can estimate a localization length $\xi$ in the scaling limit by setting $\tilde \omega \sim q^2$ and $q \approx \xi^{-1}$ and solving to obtain $\xi \sim m_0^{-2/(4-d)}$ which is finite for $d=1,2,3$ and in $d=1$ agrees with our earlier results, identifying $m_0 \leftrightarrow D_0 \Xi$.
  
  It thus follows that the isolated disordered two dimensional superconductor is described by an effective theory in which all sectors are localized with bounded single particle localization length at low energies, and with short range interactions. It then follows from \cite{BAA} that a many body localized phase should exist, at least at low energy densities, notwithstanding that the bare Hamiltonian contained long range interactions. Of course, our entire discussion is only valid in the low energy part of the spectrum, at energy densities below the `gap' scale. As such, the `rare region' scenario endemic to problems with many body mobility edges arises here also. Whether infinite temperature MBL can be obtained (or the rare region problems circumvented in some other way) is a problem that we leave to future work. 
  
We now offer an alternative, intuitive way to understand our results. A Hamiltonian with a  `Gauss law' interaction (like $\ln r$ in two spatial dimensions) can always be rewritten as a purely local Hamiltonian, with only short range interactions, by introducing a gauge field. Absent superconductivity, the obstruction to construction of a locator expansion enters in this representation through the back door, because the gauge field itself possesses an obstruction to localization \cite{GurarieChalker}, and a system where one of the sectors is protected against localization does not admit of a locator expansion \cite{QHMBL, SILL, mblbathgeneral, BanerjeeAltman}. However, once we Higgs the gauge field it loses its protection against localization, and a local Hamiltonian where none of the sectors is protected against localization can be many body localized in the usual manner. 
  
We speculate that our arguments for MBL with long range interactions may also extend to three dimensional systems with $1/r$ interactions. The basic argument follows analogously to two dimensions (the most trivial extension involves Josephson coupled superconducting layers). The gauge field now lives in three spatial dimensions, but it follows from our discussion of photon localization above that the gauge field is localized with bounded localization length at low frequency, even in $d=3$. However, there are some differences in three spatial dimensions, the full implications of which remain to be understood. One significant difference is that vortices in a three dimensional superconductor are {\it line-like} objects, and one cannot argue for their localization based simply on appeals to \cite{BAA}, which discusses localization of {\it point-like} excitations. This problem may be circumvented in one of two ways. Either one can work in the sector with no vortex excitations (easier to accomplish in three dimensions since vortex-antivortex pairs cost an energy that diverges linearly with the length of the vortex line).  Alternatively, one can appeal to the body of work establishing existence of a glassy phase of vortices that persists up to non-zero temperature in thermodynamic equilibrium \cite{Fisher, FisherFisherHuse}. If a `vortex glass' phase exists at finite temperature in thermodynamic equilibrium, then it seems plausible that a localized phase of vortices should also exist at finite energy density in an isolated quantum system.

Another point to note is that in three dimensions, the problems associated with being in the continuum (even with short range interactions) are much more severe, since {\it delocalized} single particle states arise above a critical energy. These delocalized single particle states are difficult to reconcile with MBL (although see \cite{AtoZ, Modak, SDS, MukerjeeReview}). The canonical way to regulate problems arising at high energies in the continuum is to place the theory on a lattice. In a lattice gauge theory, where the photons are also placed on a lattice, one has simply a theory of $Z_2$ topological order \cite{HanssonOganesyanSondhi}, which can be many body  localized \cite{LPQO} in the usual manner. However, for physical superconductors, the electrons live on a lattice but the gauge field lives in the continuum. As such, delocalized photon modes unavoidably appear at high energies. However, high energy photons are also non-interacting, because Maxwell's equations in vacuum are linear. Whether such non-interacting but delocalized high energy photon modes endow the system with a finite relaxation time, and how long the relaxation time is if so, is a subtlety that remains to be understood. A detailed investigation of these issues is left to future work.

 \section{Discussion}
 We have established that systems with long range interactions {\it can} be in a many body localized phase. The basic idea is that the long range interaction can drive the system into a correlated phase naturally described in terms of emergent excitations with only short range interactions, which can be many body localized in the usual manner. We have demonstrated this for a one dimensional problem of fermions with $\sim r$ interactions, and for a two dimensional problem of fermions with $\log r$ repulsion, and speculated that similar arguments may also apply to fermions in three dimensions with $\sim 1/r$ repulsion i.e. many body localization is compatible with Coulomb repulsion in all physical dimensions. Our arguments lean on \cite{BAA} and are thus `asymptotic' in the sense that they only establish localization at low temperatures. Whether infinite temperature MBL can be realized in long range interacting systems is a problem that we leave to future work. 
  
Our work brings into focus a host of additional conceptual questions. For example, does low temperature MBL have a description in terms of emergent local integrals of motion, similar to infinite temperature MBL? Recent work \cite{GBN} has provided the beginnings of an answer, but much remains to be understood. Also open is the question of whether MBL can arise in mixed dimensional problems e.g. systems of fermions moving in two dimensions, but interacting via a $1/r$ potential mediated by a gauge field that lives in three dimensions. Prima-facie this seems unlikely, since disorder in two dimensions will not localize a gauge field that lives in three dimensions, and the three dimensional gauge field could then act as a `higher dimensional bath' to delocalize the system \cite{NGADP}, however, the problem deserves more careful consideration. 

A particularly interesting open question is whether MBL can arise for interactions of range intermediate between Coulomb and short range e.g. the experimentally relevant case of dipolar interactions \cite{LukinTC}? On physical grounds, one could argue that if Coulomb interactions admit of MBL, and short range interactions admit of MBL, then interactions of intermediate range should admit of MBL also. However, the particular methods we employed to establish MBL with Coulomb interactions do not readily generalize to interactions of intermediate range. In one dimension, a density-density interaction bosonizes to $V(q) q^2 \phi^2$, where $V(q)$ is the Fourier transform of the potential. The confining potential $V(r) \sim - r$ is special in that it has $V(q) \sim 1/q^2$, which produces a mass gap in the bosonic spectrum. A less long range interaction would bosonize to a term of the form $ q^{\alpha} \phi^2$, where $0<\alpha < 2$, and this would not open up a mass gap. The mass gap, we remind the reader, was important to our argument in that it produced a {\it non-compact} potential with a unique minimum, allowing us to ignore instanton events. A term like $q^{\alpha} \phi^2$ with $\alpha > 0$ would leave us with a problem of bosons with a complicated dispersion in a {\it compact} potential, which does not appear amenable to analytical solution. Similarly, two dimensional problems with $\log r$ interactions, and three dimensional problems with $1/r$ interactions, were also special in that this interaction can be mediated by a gauge field with a `natural' kinetic energy, allowing us to map the long range interacting problem to a local theory (the Abelian Higgs model) in which all sectors can be localized. Alternative power laws for the interaction will not exhibit this nice property. In particular, if the interaction is rewritten in terms of an interaction with a bosonic `gauge field,' the kinetic energy for the gauge field will not take a natural form - and a gauge field with a non-analytic `kinetic energy' may well possess its own obstructions to localization. Thus, while {\it physically} it seems plausible that interactions of intermediate range should also admit of localization, the particular methods we have employed herein do not readily generalize, and a demonstration of MBL in such systems will require fresh ideas. One possibility may be to use the generalized `Gauss' laws' that arise for higher rank gauge fields \cite{Pretko}. A detailed investigation of the possibility of MBL with intermediate range interactions is left to future work. 

Finally, it is interesting to ask what other types of correlated phase could serve as stepping stones to MBL physics in long range interacting systems, besides the confined and Higgsed phases discussed herein. Notwithstanding these open questions, however, our demonstration of low (but non-zero) temperature MBL in long range interacting systems already opens the door to investigation of MBL physics in a host of experimentally relevant systems with long range interactions. 

{\bf Acknowledgements} We acknowledge useful conversations with Victor Gurarie and Ana Maria Rey. We thank Sarang Gopalakrishnan and Ehud Altman for feedback on the manuscript. We also acknowledge Ahmed Akhtar and M.C. Banuls for an ongoing collaboration on related ideas. RMN is supported in part by the Sloan Foundation through a Sloan Research Fellowship. SLS is supported in part by the U.S. Department of Energy under grant No. DE-SC0016244

 \appendix
\section{Localization of Higgsed plasmon}
In this Appendix we generalize the calculation of \cite{GurarieChalker} to a Higgsed Goldstone mode, and demonstrate that the localization length is bounded at low energies in spatial dimensions $d=1,2,3$. 

  Lets first review the calculation of \cite{GurarieChalker}, which considers phonons (Goldstone modes) in a random medium, with the equation of motion
  \begin{equation}
  \omega^2 \phi(r) = c(r)^2 \nabla^2 \phi(r) 
  \end{equation}
  Note that the disorder enters through a term that also carries spatial derivatives. Meanwhile, the Green functions of the phonon field take the form $G(\omega, k) = \frac{1}{\omega - E_k - i/\tau}$, where the scattering time $\tau$ comes from scattering off disorder and $E_k = c k$, with $c$ being the mean phonon speed. The scattering time may be estimated from the Self Consistent Born Approximation (SCBA), whereupon one has to solve the self consistent equation 
  \begin{equation}
  1 = g(\omega) \int \frac{ k^{d-1} dk}{(\omega - E_k)^2 + 1 / \tau^2}. \label{SCBA}
  \end{equation}
  Here $\omega$ is the phonon frequency and the disorder strength $g(\omega) \sim \omega^2$, because the disorder enters in a term that involves spatial derivatives, such that the coupling to disorder vanishes at long wavelengths/low frequencies. For $E_k \sim k$ this yields $\tau^{-1} \sim \omega^{d+1}$, and a mean free path $l \sim \tau \sim \omega^{-(d+1)}$. In one dimension, the localization length is proportional to the mean free path so $\xi_{1D} \sim \omega^{-2}$ diverges as a power law at low frequency. In two dimensions weak localization theory predicts $\xi_{2D} = \exp(k l)$. Taking $k = \omega$ and $l = \omega^{-3}$ we obtain $\xi_{2D} = \exp(1/\omega^2)$ which diverges exponentially fast at low frequencies.  In three dimensions we have $l\sim \omega^{-4}$ and $k l \sim \omega^{-3}$. In three dimensions weak localization theory predicts a delocalized phase for large $k l$ i.e. the low frequency phonon states are delocalized. 
  
The analysis can be readily generalized to a Higgsed plasmon mode, to determine whether the localization length diverges close to the gap edge. Once again, self consistent Born approximation yields an expression of the form (\ref{SCBA}). However, since the disorder in Eq.\ref{AbelianHiggs} enters through a term that is independent of spatial derivatives, the disorder vertex is frequency independent at low frequency $g(\omega) \sim g$. Additionally, the dispersion is modified to $E_k \sim \Delta + k^2$ (Eq.\ref{plasmondispersion}). SCBA now predicts $
\tau^{-1} \sim (\omega-\Delta)^{(d-1)/3}$ and a mean free path $l \sim \tau^{1/2} \sim (\omega-\Delta)^{(1-d)/6}$. In one dimension, the localization length is proportional to the mean free path, which remains finite as $\omega \rightarrow \Delta$. In two and three dimension, the control parameter for weak localization theory, $ k l \sim (\omega-\Delta)^{(4-d)/6}$ is divergence free at low frequency, and thus all states remain localized, even arbitrarily close to the gap edge. Indeed,  $k l$ {\it vanishes} close to the gap edge, indicating that close to the gap edge states are in the Ioffe-Regel `strong scattering' regime where weak localization gives way to {\it strong} localization \cite{Anderson}. Additionally, in the presence of disorder there will be `Lifshitz tail' states in the gap $\omega < \Delta$, but these are expected to be localized with bounded localization length in any dimension. One thus concludes that a Higgsed plasmon mode can have all its low frequency states localized with bounded localization length. This conclusion is also consistent with the alternative argument adapted from \cite{GurarieChalker}, presented in the main text. 

 \bibliography{LRMBL}

 \end{document}